\documentclass[amssymb,preprint,preprintnumbers,nofootinbib,superscriptaddress]{revtex4-2}
\usepackage{amsmath}
\usepackage{graphicx}
\graphicspath{{Figures/}}
\usepackage{latexsym}
\usepackage{amsfonts}
\usepackage{url,hyperref}
\usepackage{bm}
\usepackage{textcomp}
\usepackage{xcolor}
\usepackage{bbm}
\usepackage{slashed}
\usepackage{caption}
\usepackage{epstopdf}
\usepackage{subcaption}
\usepackage{placeins}
\usepackage{color}
\usepackage[a4paper,left=20mm,right=20mm,top=25mm,bottom=25mm]{geometry}
\usepackage{cancel}

\newcommand{\dd}{\text{d}}

\begin{document}

\title{$q$-Equilibrium of Gas in Spacetime of Multi-horizon Black Holes}

\author{Phuwadon Chunaksorn \footnote{Email: maxwelltle@gmail.com}}
\affiliation{The Institute for Fundamental Study, Naresuan University, Phitsanulok, 65000, Thailand}

\author{Ratchaphat Nakarachinda \footnote{Email: tahpahctar\_net@hotmail.com}}
\affiliation{The Institute for Fundamental Study, Naresuan University, Phitsanulok, 65000, Thailand}

\author{Pitayuth Wongjun \footnote{Email: pitbaa@gmail.com}}
\affiliation{The Institute for Fundamental Study, Naresuan University, Phitsanulok, 65000, Thailand}

\begin{abstract}
We investigate the possibility of describing the thermal system with different temperatures for a black hole with multiple horizons. A black hole with two horizons such as the Schwarzschild-de Sitter black hole corresponds to two thermal systems with generically different temperatures. Then, it is not suitable to describe these systems with equilibrium thermodynamics corresponding to Gibbs-Boltzmann kinetic theory. In the present work, we investigate such thermal systems by using hydrostatic equilibrium thermodynamics. Assuming that the gas between the horizons obeys the Tsallis statistical mechanics, we found that it is possible to obtain the temperature gradient for the classical gas. Interestingly, the gas behaves as classical gas near the horizon and behaves like quantum gas around flat spacetime with constant temperature. As a result, the multi-horizon black holes in hydrostatic equilibrium can be in a stable configuration with the aspect of the $q$-kinetic theory.

\end{abstract}

\maketitle{}

\newpage

\section{Introduction}

The black hole is one of the curious objects predicted by Einstein’s general theory of relativity.
Classically, the black hole does not allow all particles to escape from its horizon even light. 
However, by taking the quantum effect into account, the black hole can emit radiation, called Hawking radiation, like a black body radiation with a certain temperature known as Hawking temperature \cite{hawking1975particle}.
According to the first laws of black hole thermodynamics, the black hole entropy called Bekenstein-Hawking entropy is also defined as $S_\text{BH} = A/4$ where $A$ is the surface area at the black hole's event horizon \cite{bekenstein1973black, bekenstein1974generalized, hawking1975particle}.
There have been numerous investigations of the thermodynamic properties of black holes (for reviews, see \cite{Wald:1999vt, Carlip:2014pma, Kubiznak:2016qmn} and references therein).
The significant remark is that the thermodynamic system associated with the black hole can be defined at the horizon.
If the black hole has multiple horizons, the Hawking temperatures evaluated on distinct horizons are generally different.
The multi-horizon black hole is indicated as the systems which are not in thermodynamic equilibrium.
Consequently, there should exist a temperature gradient or heat transfer between the aforementioned systems.
It is interesting to investigate the behaviors of a gas in the curved spacetime between black hole’s horizons.

In the statistical mechanic framework, the macroscopic behavior of a thermal gas system is described by motions of microscopic constituents based on the mechanical descriptions.
The collisions among fine-grained gas particles are simply treated as local interaction and also satisfy the Boltzmann Stosszahlansatz.
As a result, the kinetic behaviors of numerous gas particles can be characterized by a one-particle distribution function defined in the phase space.
The macroscopic quantities, e.g. the number of particles, energy-momentum tensor, and entropy, are determined by considering the transfers of suitable fields in the momentum space.
According to the transfer equation at equilibrium, or equivalently the balance equation, the Gibbs-Boltzmann (GB) entropy flow is proven from the well-known $H$-theorem that the entropy never decreases.

One of the important properties of the GB entropy is additive, $S = S_{(1)} + S_{(2)}$, where $S_{(1)}$ and $S_{(2)}$ are the entropy of the subsystems $1$ and $2$, respectively. 
The additivity of the entropy refers to ignoring the long-range interaction, such as gravitation. 
In addition, ignoring the long-range interaction leads to additive internal energy, $E = E_{(1)} + E_{(2)}$, where $E_{(1)}$ and $E_{(2)}$ are, respectively, the internal energy of the subsystem $1$ and $2$. 
Consequently, at thermodynamic equilibrium, the empirical temperature can be defined by utilizing the principle of the maximum entropy, $\delta S=0$ (and the energy is also minimum $\delta E=0$), as $1/T=\partial S_{(1)}/\partial E_{(1)}=\partial S_{(2)}/\partial E_{(2)}$
Such an empirical temperature is compatible with the zeroth law of thermodynamics, $T_{1} = T_{2}$, leading to the thermal equilibrium.

GB statistical mechanics has been applied to many physical phenomena, and it is always compatible with the isothermal system.
However, including the long-range interaction, it suffers from the thermodynamic instability known as Antonov's instability or gravothermal catastrophe \cite{antonov1962most, lynden1968gravo, chavanis2002gravitational, Axenides:2012bf, Roupas:2014sda}. 
This infers that GB statistical mechanics is not suitable to describe the self-gravitating system due to the long-range interaction. In fact, by taking the long-range interaction into account, the total energy and/or the total entropy lack additivity and extensivity. In order to investigate the self-gravitating system, statistical mechanics with generalized entropy should be introduced to describe such a system rather than GB statistical mechanics.

One of the generalized entropies is Tsallis entropy \cite{tsallis1988possible} obeying the pseudo-additive composition rule, $S = S_{(1)} + S_{(2)} + (1-q) S_{(1)} S_{(2)}$ \cite{Curado:1991jc, abe2001nonextensive}, where $q > 0$ is a non-extensive parameter. 
With the pseudo-additivity, the empirical temperature obtained from the principle of the maximum entropy is not clear how to define. Consequently, the empirical temperature obtained from Tsallis entropy is not compatible with the zeroth law of thermodynamics~\cite{biro2011zeroth}, leading to the thermodynamic in-equilibrium. As a result, Tsallis statistical mechanics has been widely investigated in non-isothermal self-gravitating systems. For example, the Tsallis entropy was applied to the stellar polytropes of the self-gravitating systems to seek and discuss their stability \cite{taruya2002gravothermal, taruya2003gravothermal, Taruya:2002jz, Sakagami:2003qs, Taruya:2005bz}. With Tsallis statistical mechanics, the $H$-theorem, and the Boltzmann transport equation were derived \cite{lima2001nonextensive, lavagno2002relativistic, silva2005relativity, Santos:2017gzy}. The relation between the gradient of the temperature and the gradient of the gravitational potential for the self-gravitating systems is obtained, and, using these results, the thermodynamic behaviors of the spatial in-homogeneous temperature are analyzed \cite{lima2001nonextensive, jiulin2004nonextensive, jiulin2007nonextensivity,liu2011nonextensivity, zheng2017limit}. Besides the investigations as mentioned earlier, statistical mechanics, associated with Tsallis entropy, is also applied to cosmology \cite{sheykhi2018modified, Saridakis:2018unr, Ghoshal:2021ief, Lymperis:2018iuz, Jizba:2023fkp, 2020EPJC...80...25S, Dehpour:2023dfo}. In particular, it has been argued that the modified Friedmann equation admits the accelerated expansion without appealing to any kind of dark energy models. It was pointed out that the non-extensivity in the context of black hole thermodynamics can infer the generalized uncertainty principle \cite{PhysRevD.103.026021, Cimidiker:2023kle}. Regarding Bekenstein-Hawking entropy as Tsallis one, the thermodynamic properties and the stability of the black holes were analyzed \cite{Biro:2013cra, Czinner:2015eyk, Czinner:2017tjq, Tannukij:2020njz, Promsiri:2020jga, Promsiri:2021hhv,  Nakarachinda:2021jxd, Hirunsirisawat:2022fsb, Chunaksorn:2022whl, Nakarachinda:2022gsb}. Furthermore, there have been generalized entropies employed to investigate physical systems such as black hole thermodynamics and cosmology \cite{Nojiri:2022aof, Nojiri:2022dkr, Nojiri:2022sfd, Nojiri:2022ljp, Odintsov:2022qnn}.
\par
Although there are various statistical mechanics approaches dealing with long-range interaction,  the relativistic kinetic theory turns out to be the most efficient and convenient option. In fact, it is useful to apply to various contexts ranging from stability theory \cite{1994JMP....35.4809B, Fajman:2017gaz, Rein:2023iic},
astrophysics \cite{Rioseco:2016jwc, Cieslik:2022wok} to cosmology \cite{Weinberg:2003ur, Agon:2011mz}. Moreover, we aim to study gaseous systems in curved spacetime compatible with relativistic systems. Then it is suitable to use the relativistic kinetic theory in the present work. There have been firstly developed and generalized the non-relativistic distribution functions for the classical gas and quantum gas to the relativistic ones by J\"uttner \cite{Juttner1911, Juttner1928}. Particularly, the equation of state describing such a case is also derived. The covariant kinetic theory of relativistic gases has
been constructed by Synge who studied a system of gases \cite{Synge:1934zzb}. 
\par
In curved spacetime, Tolman and Ehrenfest found that relativistic gases can only be in equilibrium when the effects of the temperature gradient and the gravitational field cancel out \cite{PhysRev.35.904, PhysRev.36.1791}, and the kinetic theory on curved spacetime was formulated by Tauber and Weinberg \cite{Tauber:1961lbq}. The hydrodynamic equations for fluid on curved spacetime were derived from the Boltzmann transport equation by using the Chapman-Enskog method \cite{1963JMP.....4.1163I}, and then applied such equations to the gaseous system composed of zero rest mass particles in spherically symmetric spacetime \cite{1966AnPhy..37..487L}. Fourier’s laws for a single gas and Fick’s law for gas mixtures were obtained by considering the gases in Schwarzschild (Sch) spacetime under
the post-Newton approximation \cite{Kremer:2012nk}. For the kinetic theory in the language of modern differential geometry, there are investigations such as the geometric formulation on the tangent bundle \cite{Sarbach:2013fya, Sarbach:2013uba, Sarbach:2013hna} and the cotangent bundle \cite{Acuna-Cardenas:2021nkj}.
\par
In this work, we investigate the possibility of describing the thermal systems of a black hole with multiple horizons. 
We determine such thermal systems in the presence of a gravitational field \cite{Walker_1936, cercignani2002relativistic} based on the $q$-kinetic theory \cite{lima2001nonextensive, lavagno2002relativistic, silva2005relativity, Santos:2017gzy,jiulin2004nonextensive, jiulin2007nonextensivity,liu2011nonextensivity, zheng2017limit}.
By assuming that the gas between the horizons obeys the Tsallis statistical mechanics, it is found that there exists the temperature gradient for the classical gas, but it does not for the quantum gas. In fact, the temperature gradient depends on the gradient of the chemical potential. Due to the similar roles of the chemical potential and gravitational potential, we set them to be proportional to each other in order to characterize the behavior of the gases between black hole's horizons. Interestingly, the gas near the black hole horizon behaves as a classical one, while it will be observed the quantum nature of the gas far away from the black hole with the same temperature as that from Hawking radiation. We also found that a similar situation can be obtained for a black hole with a single horizon such as the Sch black hole as well as one with an inner horizon such as a charged black hole. As a result, the multi-horizon black holes in hydrostatic equilibrium can be in a stable configuration with the aspect of the $q$-kinetic theory.
\par
This work is organized as follows. In Sec.~\ref{sec:GB}, the relativistic Boltzmann transport equation based on GB entropy in the gravitational field is reviewed. 
The sufficient and necessary conditions thermodynamic equilibrium of the systems are presented. It is found that GB statistical mechanics provides the self-gravitating system being always isothermal. 
In Sec.~\ref{sec:q-stat}, the Tsallis entropy and the corresponding $q$-Boltzmann transport equation are introduced. 
We construct the systems' $q$-power law distribution function, and the expressions of physical quantities for the non-extensive parameter, $q\neq1$, are derived. We will see from the obtained results that the temperature gradient of the systems does not vanish, specifying a non-isothermal configuration. 
With those thermodynamic quantities, they have been adopted to construct the thermodynamic quantities of the static and spherically symmetric black holes in Sec.~\ref{sec:bh}, 
In Sec.~\ref{sec:conclude}, the main conclusion and discussion of this work are stated. 
Throughout this paper, the reduced Planck constant $\hbar$, Boltzmann constant $k_\text{B}$, speed of light $c$, and Newton’s gravitational constant $G$ are set as unity for convenience.

\section{The relativistic Boltzmann transport equation}\label{sec:GB}

This section is devoted to recapping the key results of the kinetic theory of the relativistic gas in the presence of the gravitational field. 
Let us start with considering relativistic particles of a single non-degenerate gas, i.e., the classical gas. 
Each particle can be characterized by the spacetime coordinates $x^{\alpha}$ 
and the four-momentum $P^\alpha = (P^0, P^i)$. The (curved) spacetime is described by metric $g_{\alpha\beta}$. 
The states of the particle in the phase space $(x^\alpha, P^i)$ is kinetically characterized by the one-particle distribution function, $\rho(x^\alpha, P^\alpha)= \rho(x^\alpha, P^i)$. Note that the temporal component of the four-momentum depends on the spatial ones obeying the constraint: $P^\alpha P_\alpha=-m^2$. For the particles with the momentum $\dd^{3}P/P_0$ passing through a hypersurface $n_\alpha\dd\Sigma$ where $n^\alpha$ is the normal vector of the hypersurface, the change in the number of particles is defined as $\dd N=\rho(x^\mu, P^i)\dd\Gamma$ where the invariant volume element of the phase space is given by $\dd\Gamma=P^\alpha n_\alpha\dd\Sigma\dd^{3}P/P_0$. The collisions of the gas are assumed to satisfy the Boltzmann Stosszahlansatz which is composed of three following assumptions. 1) the gas is sufficiently dilute so that only the collisions of pairs of particles are considered. 2) the assumption of molecular chaos: pairs of incoming particles (before collision) are not correlated. The pairs of outgoing ones (after collision) are also obeyed. 3) the variation of the one-particle distribution function over time interval is sufficiently slow. Such a time interval is much shorter than the time between two collisions, but much longer than the duration of a collision. As a result, the net rate of change in the number of the particles due to the particles' local binary collision yields the time evolution of the one-particle distribution function known as the classical Boltzmann transport equation \cite{Walker_1936, cercignani2002relativistic}
\begin{eqnarray}
\displaystyle
P^\alpha\frac{\partial \rho}{\partial x^\alpha}-\Gamma^\gamma_{\alpha\beta}P^\alpha P^\beta\frac{\partial \rho}{\partial P^\gamma} = C[\rho],\label{Boltz eq}
\end{eqnarray}
where $\Gamma^\gamma_{\alpha\beta}$ is the Levi-Civita connection.
The left-hand side is, in fact, the total derivative of the distribution function with respect to the particles' proper time due to the particles' local binary collision denoted by $C[\rho]$ on the right-hand side.
$C[\rho]$ is the collision functional given by
\begin{eqnarray}
\displaystyle
C[\rho]=\int\frac{\dd^3P_*}{P_{*0}}\sqrt{-g}\big(\rho'_*\rho'-\rho_*\rho\big)F\sigma\dd\Omega.
\end{eqnarray}
Here, $F$, $\sigma$, $\dd\Omega$, and $g$ are the invariant flux, the differential cross-section of the collision, the element of the solid angle, and the determinant of the metric tensor, respectively. Due to the curvature of spacetime, the integral is that over the invariant element of the momentum space, $\sqrt{-g}\,\dd^3P_*/P_{*0}$. We also have used the abbreviations such that the distribution functions for the incoming particles (before the collision) with the momenta $P^i$ and $P^i_*$ are $\rho=\rho(x^\alpha, P^i)$ and $\rho_*=\rho(x^\alpha, P^i_*)$, respectively. Similarly, the distribution functions $\rho'=\rho(x^\alpha, P'^i)$ and $\rho'_*=\rho(x^\alpha, P'^i_*)$ stand for those of the outgoing particles (after the collision) with the momenta $P'^i$ and $P'^i_*$, respectively. It is very important to emphasize that, in the Boltzmann transport equation~\eqref{Boltz eq}, the contribution from external forces is ignored.
\par
The gaseous system is said to be in equilibrium when the collision term $C[\rho]$ in Eq.~\eqref{Boltz eq} vanishes.
At the end of this section, it will be seen that such a condition corresponds to the equilibrium associated with no net change in entropy four-flow.
The equilibrium distribution function denoted by that with superscript $``(0)"$, therefore, satisfies a condition
\begin{eqnarray}
\displaystyle
\rho'^{(0)}_*\rho'^{(0)}-\rho^{(0)}_*\rho^{(0)}=0,\label{equil GB 1}
\end{eqnarray}
or, equivalently,
\begin{eqnarray}
\displaystyle
\ln\rho'^{(0)}_*+\ln\rho'^{(0)}=\ln\rho^{(0)}_*+\ln\rho^{(0)}.\label{equil GB 2}
\end{eqnarray}
This is a summational invariant which has been proved that the distribution function must be taken in the form of \cite{cercignani2002relativistic}
\begin{eqnarray}
\displaystyle
\ln\rho^{(0)}(x^\alpha, P^\alpha)
\sim A(x^\alpha)+B_\beta(x^\alpha) P^\beta,\label{equi dist fn gen}
\end{eqnarray}
where $A(x^\alpha)$ and $B_\beta(x^\alpha)$ are arbitrary scalar and vector, respectively. 
It is noticed that the equilibrium distribution function $\rho^{(0)}(x^\alpha, P^\alpha)$ is satisfied the on-shell condition: $P^\alpha P_\alpha=-m^2$.
A useful form of $\rho^{(0)}$ has been explored and well-known as the Maxwell-J\"uttner distribution function \cite{Chernikov1964}
\begin{eqnarray}
\displaystyle
\rho^{(0)}(x^\alpha, P^\alpha)
=K(x^\alpha)\exp\left[\frac{\mu(x^\alpha)-v_\beta P^\beta}{T(x^\alpha)}\right],\label{MJ dist fn}
\end{eqnarray}
where $\mu(x^\alpha)$, $T(x^\alpha)$, and $v^\beta$ are the chemical potential, temperature, and four-velocity of the gas particles, respectively. 
$K(x^\alpha)$ is just a normalization factor. 
In fact, the above form of the distribution function is obtained by identifying the arbitrary functions in Eq.~\eqref{equi dist fn gen} as $A=\mu/T$ and $B^\alpha=-v^\alpha/T$. 
\par
By substituting the distribution function in Eq.~\eqref{MJ dist fn} into the Boltzmann transport equation~\eqref{Boltz eq}, one straightforwardly obtains
\begin{eqnarray}
\displaystyle
P^\alpha\left[\partial_\alpha K+K\partial_\alpha\left(\frac{\mu}{T}\right)\right]
-\frac{1}{2}P^\alpha P^\beta K\left[\nabla_\alpha\left(\frac{v_\beta}{T}\right)+\nabla_\beta\left(\frac{v_\alpha}{T}\right)\right]=0.
\end{eqnarray}
Since, on the left-hand side of the above equation, the terms containing different orders of the momentum are linearly independent, both the terms in the square bracket must be both zero. Let us focus on the quadratic term of the momentum, i.e., $\nabla_\alpha\big(v_\beta/T\big)+\nabla_\beta\big(v_\alpha/T\big)=0$.
It is further written as
\begin{eqnarray}
\displaystyle
T\big(\nabla_\alpha v_\beta+\nabla_\beta v_\alpha\big)
=v_\alpha\partial_\beta T+v_\beta\partial_\alpha T.\label{p^2 term}
\end{eqnarray}
Contracting Eq.~\eqref{p^2 term} with $v^\alpha v^\beta$, and then employing the property $v_\alpha v^\alpha=-1$ and $v_\alpha\nabla_\beta v^\alpha=0$, the condition on the gradient of the temperature is given by 
\begin{eqnarray}
\displaystyle
v^\alpha\partial_\alpha T=0.\label{grad T}
\end{eqnarray}
There is no temperature gradient of the gas in the direction of the flow. In addition, contracting Eq.~\eqref{p^2 term} with $v^\alpha$, the other thermodynamic relations can be obtained as
\begin{eqnarray}
\displaystyle
v^\alpha\nabla_\alpha v_\beta = -\frac{1}{T}\partial_\beta T.
\label{geodesic like}
\end{eqnarray}
It is noticed that the left-hand side refers to the parallel transport of the four-velocity which vanishes if the gas particles move along the geodesic in the curved spacetime. Therefore, the temperature gradient on the right-hand side can be interpreted as an effective external force that governs the particles' trajectory deviating from the geodesic. This implies that the system of gas in this spacetime can form due to the presence of the temperature gradient which is interpreted as the pressure counter-balancing the gravitational collapse.
\par
For the quantum gas system, the assumption of molecular chaos for the scattering; $P^{i} + P^{i}_{*} \rightarrow P^{'i} + P^{'i}_{*}$, is still held. However, the collision term is modified by the factor $(1 + \xi \rho_{(Q)})$ where $\xi = 1$ for the bosonic gas and $\xi = -1$ for the fermionic gas. As a result, the collision functional is written as
\begin{eqnarray}
\displaystyle
C_{\text{(Q)}}[\rho] = \int \frac{\dd^{3} P_{*}}{P_{*0}} \sqrt{-g} \Big[\rho'_{*(Q)}\rho'_{(Q)} \left(1 + \xi \rho_{*(Q)}\right) \left(1 + \xi \rho_{(Q)}\right)&& \nonumber\\
&&\hspace{-6cm}- \rho_{*(Q)} \rho_{(Q)} \left(1 + \xi \rho'_{*(Q)}\right) \left(1 + \xi \rho'_{(Q)}\right)\Big] F \sigma \dd\Omega.
\end{eqnarray}
At the thermodynamic equilibrium, the equilibrium distribution function satisfies the condition as
\begin{equation}
\displaystyle
\frac{\rho'^{(0)}_{*(Q)} \rho'^{(0)}_{(Q)}}{\left(1 + \xi \rho'^{(0)}_{*(Q)}\right) \left(1 + \xi \rho'^{(0)}_{(Q)}\right)} = \frac{\rho^{(0)}_{*(Q)} \rho^{(0)}_{(Q)}}{\left(1 + \xi \rho^{(0)}_{*(Q)}\right)\left(1 + \xi \rho^{(0)}_{(Q)}\right)},
\end{equation}
or, equivalently
\begin{equation}
\displaystyle
\ln \left(\frac{\rho'^{(0)}_{*(Q)}}{1 + \xi \rho'^{(0)}_{*(Q)}}\right) + \ln \left(\frac{\rho'^{(0)}_{(Q)}}{1 + \xi \rho'^{(0)}_{(Q)}}\right) = \ln \left(\frac{\rho^{(0)}_{*(Q)}}{1 + \xi \rho^{0}_{*(Q)}}\right) + \ln \left(\frac{\rho^{(0)}_{(Q)}}{1 + \xi \rho^{(0)}_{(Q)}}\right).
\end{equation}
The above equation is the summational invariant, and hence the distribution function can be written, in terms of conserved quantity $P^{\alpha}$, as
\begin{equation}
\displaystyle
\rho^{(0)}_{(Q)} (x^{\alpha}, P^{\alpha}) = \frac{1}{\exp \displaystyle \left[\frac{v_{\alpha} P^{\alpha} - \mu(x^{\beta})}{T_{(Q)}(x^{\beta})}\right] - \xi}.
\label{quan dist fn}
\end{equation} Substituting Eq. (\ref{quan dist fn}) into Eq. (\ref{Boltz eq}), one obtains
\begin{equation}
\displaystyle
P^{\alpha} \partial_{\alpha} \left(\frac{\mu}{T_{(Q)}}\right) - \Gamma_{\alpha \beta}^{\gamma} P^{\alpha} P^{\beta} \left[\nabla_{\alpha} \left(\frac{v_{\beta}}{T_{(Q)}}\right) + \nabla_{\beta} \left(\frac{v_{\alpha}}{T_{(Q)}}\right)\right] = 0. \label{quantum 1}
\end{equation}
By following the same manners of calculation as performed in the case of classical gas, we obtain
\begin{equation}
\displaystyle
v^{\alpha} \partial_{\alpha} T_{(Q)} = 0,\label{grad T quant}
\end{equation}
which is the same as the condition for the classical gas. Furthermore, one also has the following equation as
\begin{eqnarray}
\displaystyle
v^\alpha\nabla_\alpha v_\beta = -\frac{1}{T_{(Q)}}\partial_\beta T_{(Q)}.
\label{geodesic like quantum}
\end{eqnarray}
\par
It is important to note that, for a static and spherically symmetric gaseous system. The temperature depends only on the radial coordinate: $T=T(r)$.
The non-trivial conditions of Eqs.~\eqref{grad T} and \eqref{geodesic like}, respectively, read 
\begin{eqnarray}
\partial_r T(r)=0,
\qquad
\partial_r T_{(Q)} (r)=0.
\end{eqnarray}
In addition, Eq.~\eqref{geodesic like} becomes the geodesic equation without the effective external force due to temperature gradient.
Hence, as mentioned so far, the system of the classical and quantum gas, assumed to be described by GB statistical mechanics, is isothermal and unstable due to gravitational collapse.
\par
Now, it is useful to turn our attention to discussing how the equilibrium can be evaluated via the vanishing divergence of entropy four-flow.  
To realize this, let us consider the flow of an arbitrary scalar quantity $\psi(x^\alpha, P^i)$ along the flow of the gas particles.
For the classical gas, the dynamics of such a flow can be described by the transfer equation expressed as 
\cite{cercignani2002relativistic}
\begin{eqnarray}
\displaystyle
\nabla_\alpha\left[\int\frac{\dd^3P}{P_0}\sqrt{-g}P^\alpha\rho\,\psi\right]
+\int\frac{\dd^3P}{P_0}\sqrt{-g}P^\alpha\rho
\left(-\partial_\alpha\psi+\Gamma^i_{\alpha\beta}P^\beta\frac{\partial\psi}{\partial P^i}\right)&&\nonumber\\
&&\hspace{-10cm}=\frac{1}{4}\int\frac{\dd^3P_*}{P_{*0}}\frac{\dd^3P}{P_0}(-g)\big(\psi+\psi_*-\psi'-\psi'_*\big)\big(\rho'_*\rho'-\rho_*\rho\big)F\sigma\dd\Omega.\label{transf eq}
\end{eqnarray}
The right-hand side refers to the production term due to collisions.
One of the important consequences of this equation is that the conserved quantities along the gas flow can be obtained by identifying the proper form of the scalar function.
For example, by choosing $\psi=1$, the transfer equation~\eqref{transf eq} yields the conservation of the number of gas particles $\nabla_\alpha N^\alpha=0$, where the particle four-flow is defined by $\displaystyle N^\alpha=\int\dd^3P\sqrt{-g}P^\alpha\rho/P_0$.
It is worthwhile to note that the macroscopic description of the gas can be determined by the transfer equation. 
Now, let us investigate the dynamics of the entropy four-flow.
The GB entropy can be expressed in terms of the distribution function $\rho$ as follows: 
\begin{equation}
\displaystyle
S_\text{GB}=-\int\dd\Gamma\rho\ln\rho.\label{SGB}
\end{equation}
Its flow is simply defined by integrating the integrand in Eq.~\eqref{SGB} along the momentum of the gas over the momentum space, i.e., 
\begin{eqnarray}
\displaystyle
S^\alpha=-\int\frac{\dd^3P}{P_0}\sqrt{-g}P^\alpha\rho\ln\rho.\label{S 4 flow GB}
\end{eqnarray}
As a result, the dynamics of the entropy four-flow governed by Eq.~\eqref{transf eq} is obtained by identifying the arbitrary scalar function as $\psi=-\ln\rho$.
Hence, Eq.~\eqref{transf eq} can be rewritten as
\begin{equation}
\displaystyle
\nabla_{\alpha} S^{\alpha} = \frac{1}{4} \int \frac{\dd^{3} P_{*}}{P_{*0}} \frac{\dd^{3}P}{P_{0}} (-g) \rho_{*} \rho \left[\left(\frac{\rho^{'}_{*} \rho^{'}}{\rho_{*} \rho} - 1\right) \ln \left(\frac{\rho^{'}_{*} \rho^{'}}{\rho_{*} \rho}\right)\right] F \sigma \dd\Omega.
\label{H-theorem GB}
\end{equation}
Note that the second term on the left-hand side of Eq.~\eqref{transf eq} vanishes by using Eq.~(\ref{Boltz eq}) and using the fact that the phase space volume element is invariant. 
It is observed that the term in the square bracket is taken in the form of $(x - 1) \ln x$ which is non-negative, $(x - 1) \ln x \geq 0$ where $x\equiv\frac{\rho'_*\rho'}{\rho_*\rho}>0$. 
In addition, it is also found that $(x-1) \ln x = 0$ as $x=1$.
Accordingly, the dynamics of the entropy four-flow is expressed as
\begin{equation}
\displaystyle
\nabla_{\alpha} S^{\alpha} \geq 0,\label{H-theorem 2 GB}
\end{equation}
which is the well-known $H$-theorem for the entropy four-flow $S^\alpha$. This result remarkably corresponds to the second law of thermodynamics.
For the scenario that $\nabla_{\alpha} S^{\alpha} = 0$, this is equivalent to the system without the net entropy flow. As a result, it is worthwhile to define the equilibrium state satisfying $\nabla_{\alpha} S^{\alpha} = 0$.
By using Eq.~\eqref{H-theorem GB}, the conditions for the equilibrium state can be written in terms of the distribution function as mentioned in Eqs.~\eqref{equil GB 1} and~\eqref{equil GB 2}.
Let us emphasize that the equilibrium state implies the vanishing of the collision term $C[\rho]$. 
\par
Before ending this section, it is worthwhile to summarize the key results as follows.
By treating the system of (classical/quantum) gas in the gravitational field to obey GB statistical mechanics, it is found that the gas is always isothermal and cannot be in a stable configuration with the aspect of the kinetic theory.
In order to obtain the statistical description of the non-isothermal system, it is worthwhile to consider other statistical mechanics instead of GB ones. 
A possible way is to modify the transfer equation~\eqref{transf eq}.
Accordingly, we reviewed the basic knowledge of how to define the equilibrium from the transfer equation in the last part of this section.
In the next section, one of the generalized entropies known as non-extensive Tsallis entropy \cite{tsallis1988possible, Curado:1991jc, gell2004nonextensive} will be introduced. 
The kinetic theory for such an entropy is known as $q$-kinetic theory, a one parameter-extended theory of GB one.

\section{The kinetic theory based on Tsallis statistical mechanics} \label{sec:q-stat}
\par
In this section, the thermodynamic quantities of the $N$-relativistic particles in the presence of the gravitational fields are investigated by using statistical mechanics based on Tsallis entropy. 
The Tsallis entropy with one non-extensive parameter, $q>0$, can be defined in terms of the one-particle distribution function as follows: \cite{tsallis1988possible, Curado:1991jc, gell2004nonextensive}
\begin{equation}
\displaystyle
S_q=-\int\dd\Gamma\rho^q\ln_q\rho.\label{Sq}
\end{equation}
where $\ln_q\rho$ is the $q$-logarithmic function of the distribution function.
Such a logarithm and its inverse (i.e., $q$-exponential function) of an arbitrary positive function $X$ can be, respectively, expressed as
\begin{eqnarray}
\displaystyle
\ln_qX&=&\frac{1}{1-q}\left(X^{1-q}-1\right)\label{q-ln},\\ 
\exp_qX&=&\big[1+(1-q)X\big]^{\frac{1}{1-q}}.
\end{eqnarray}
One of the significant features of the Tsallis entropy is its non-extensivity which is obtained from the composition identity of the $q$-logarithmic function
\begin{eqnarray}
\displaystyle
\ln_qX+\ln_qY
&=&\ln_q\big(XY\big)-(1-q)\ln_qX\ln_qY.
\end{eqnarray}
It is noticed that the last term on the right-hand side disappears when $q=1$.
This property implies the pseudo-additive composition rule of the Tsallis entropy: $S_{q(1+2)}=S_{q(1)}+S_{q(2)}+(1-q)\,S_{q(1)}S_{q(2)}$, where the subscript $(1+2)$ denotes the quantity of the combined system 1 and 2 while those of subsystems are denoted by $(1)$ and $(2)$. In addition, the Tsallis entropy in Eq.~\eqref{Sq} can recover the GB one by taking the limit of the non-extensive parameter approaching unity, i.e., $\displaystyle{\lim_{q\to1}S_q=S_\text{GB}}$. The composition rule also reduces to the additive one in this limit.
\par
For the kinetic behavior of the non-extensive systems, such as self-gravitating systems, the relativistic gas can be characterized by $q$-distribution function, $\rho_q (x^\alpha, P^\alpha)$, in which its time evolution obeys the $q$-Boltzmann transport equation in the presence of the gravitational field as \cite{lima2001nonextensive, lavagno2002relativistic, Santos:2017gzy}
\begin{equation}
\displaystyle
P^\alpha\frac{\partial\rho_q}{\partial x^\alpha}-\Gamma^\alpha_{\beta\gamma}P^\beta P^\gamma\frac{\partial\rho_q}{\partial P^\alpha}
=C_q[\rho_q].\label{Boltz eq q}
\end{equation} 
Note that, by comparing to Eq.~\eqref{Boltz eq}, the non-extensivity does not affect the form of the left-hand side of Eq.~\eqref{Boltz eq q} because the mentioned terms are relevant only to the geometry of spacetime, not to the properties of particles' collisions.
The non-extensivity effect is, thus, incorporated through the $q$-generalized collision term:
\begin{equation}
\displaystyle
C_q[\rho_q]=\int\frac{\dd^3P_*}{P_{*0}}\sqrt{-g}R_q(\rho_q, \rho'_q)F\sigma\dd\Omega,
\label{Cq}
\end{equation}
where $R_q(\rho_q, \rho'_q)$ is the difference of correlation functions associated with before- and after-colliding gas particles.
As proposed in Ref.~\cite{lima2001nonextensive}, the form of $R_{q} (\rho_q,\rho_q')$ reads
\begin{eqnarray}
\displaystyle
R_q(\rho_q, \rho'_q) 
&=&\exp_q\Big[\big(\rho'_{q\,*}\big)^{q-1}\ln_q\rho'_{q\,*}+\big(\rho'_q\big)^{q-1}\ln_q\rho'_q\Big]\nonumber\\
&&-\exp_q\Big[\big(\rho_{q\,*}\big)^{q-1}\ln_q\rho_{q\,*}
+\big(\rho_q\big)^{q-1}\ln_q\rho_q\Big].
\label{Rq}
\end{eqnarray}
It is important to emphasize that the collision term $C_q[\rho_q]$ with above $R_q(\rho_q, \rho'_q)$ is assumed to satisfy the $q$-generalized form of the molecular chaos hypothesis rather than the traditional version.
In other words, the recent collision term cannot satisfy Boltzmann's Stosszahlansatz, and the assumption of the collision of the gas described by $q$-statistical mechanics is then necessarily modified \cite{lima2001nonextensive, silva2005relativity, Abe2009}.
At the GB limit, one can check that $\displaystyle{\lim_{q\to1}R_q(\rho_q,\rho'_q)=(\rho'_*\rho'-\rho_*\rho})$ or, equivalently, $\displaystyle{\lim_{q\to1}C_q[\rho_q]=C[\rho]}$ as expected. Straightforwardly, the transfer of the Tsallis entropy is given by
\begin{eqnarray}
\displaystyle
S_q^\alpha=-\int\frac{\dd^3P}{P_0}\sqrt{-g}P^\alpha\rho^q\ln_q\rho.
\end{eqnarray}
It is found that there exists the inequality condition $\nabla_\alpha S_q^\alpha\geq0$ which is associated with the $q$-generalization of the $H$-theorem \cite{lima2001nonextensive, silva2005relativity}.
\par
The equilibrium $q$-distribution function is defined as a function satisfying the equilibrium condition: $C_q[\rho_q]=0$.
From the collision in Eq.~\eqref{Cq} together with Eq.~\eqref{Rq}, the equilibrium $q$-distribution function has to obey the summational invariant:
\begin{eqnarray}
\displaystyle	\big(\rho'^{(0)}_{q\,*}\big)^{q-1}\ln_q\rho'^{(0)}_{q\,*}+\big(\rho'^{(0)}_q\big)^{q-1}\ln_q\rho'^{(0)}_q
=\big(\rho^{(0)}_{q\,*}\big)^{q-1}\ln_q\rho^{(0)}_{q\,*}+\big(\rho^{(0)}_q\big)^{q-1}\ln_q\rho^{(0)}_q.
\end{eqnarray}
Similar to the argument in the previous section, one then obtains
\begin{eqnarray}
\displaystyle	\big(\rho^{(0)}_q\big)^{q-1}\ln_q\rho^{(0)}_q
\sim A_q(x^\alpha)+(B_q)^\beta(x^\alpha) P_\beta,
\end{eqnarray}
where $A_q$ and $(B_q)^\beta$ are the coefficients associated with the aforementioned summational invariant.
By a few steps of algebraic rearrangement, the equilibrium $q$-distribution function can be expressed as
\begin{eqnarray}
\displaystyle
\rho^{(0)}_q(x^\alpha,P^\alpha)
&=&K_q(x^\alpha)\bigg[1-(1-q)\Big\{A_q(x^\alpha)+(B_q)^\beta(x^\alpha) P_\beta\Big\}\bigg]^{-1/(1-q)},
\end{eqnarray}
where $K_q(x^\alpha)$ is the normalization factor. It is found that the above equilibrium distribution function with $q\to1$ is reduced to the GB one in Eq.~\eqref{MJ dist fn} when the coefficients are set by $K_{q=1}=K$, $A_{q=1}=A$ and $\big(B_{q=1}\big)_\beta=B_\beta$ as expected. Plugging this $q$-distribution function in the $q$-Boltzmann transport equation~\eqref{Boltz eq q}, the vanishing conditions of both linear and quadratic terms in momentum, respectively, yield
\begin{eqnarray}
\displaystyle
\partial_\alpha K_q\big[1-(1-q)A_q\big]+K_q\partial_\alpha A_q&=&0,\\
\partial_\alpha K_q\Big[-(1-q)\big(B_q\big)_\beta\Big]+\frac{K_q}{2}\Big[\nabla_\alpha\big(B_q\big)_\beta+\nabla_\beta\big(B_q\big)_\alpha\Big]&=&0.
\end{eqnarray}
It is possible to solve the above system for a non-trivial condition of $A_q$ and $(B_q\big)_\beta$ as follows:
\begin{eqnarray}
\displaystyle
-2(1-q)\partial_\alpha A_q\big(B_q\big)_\beta
=\big[1-(1-q)A_q\big]\Big[\nabla_\alpha\big(B_q\big)_\beta+\nabla_\beta\big(B_q\big)_\alpha\Big].
\label{constraint q}
\end{eqnarray}
In the same analysis fashion, the coefficients are interpreted as the standard fields so that $A_q=\mu/T_q$ and $(B_q\big)_\beta=v_\beta/T_q$. First contracting $v^\alpha v^\beta$ to Eq.~\eqref{constraint q} with the aforementioned substitution and also using the identities: $v_\alpha v^\alpha=-1$ and $v^\alpha\nabla_\beta v_\alpha=0$.
As a result, we obtain
\begin{eqnarray}
\displaystyle
v^\alpha\partial_\alpha T_q
=(1-q)v^\alpha\partial_\alpha\mu.
\label{grad T q}
\end{eqnarray}
Obviously, without the non-extensivity (i.e., $q=1$), Eq. (\ref{grad T q}) can recover to Eq.~\eqref{grad T}. For the q-generalization, the temperature gradient of the non-extensive
system exists due to the particle transfer characterized by the gradient of the chemical potential.
\par
Considering the self-gravitation system, gravity radially attracts all surrounding gas particles to form the sphere of the gas. The particles in the system move from the larger radius to the smaller one. This feature is characterized by the gravitational potential $\Phi(r)\propto-1/r$, where the negative sign refers to the attractive interaction.
Therefore, in this sense, the gas particles transfer from a higher gravitational potential to a lower one. In the thermodynamic framework, the transfer is indeed characterized by the chemical potential $\mu$. Furthermore, the sign of the chemical potential for a classical system is negative. To illustrate it, let us consider, for example, the system of two distinguishable particles at thermodynamic equilibrium with the internal energy, $E=2\epsilon$. A number of the microstates for arranging these particles in all possible energy levels can be counted as three states so that the possible microstates are $\{(0,2\epsilon),(2\epsilon,0),(\epsilon,\epsilon)\}$. Therefore, the entropy of the system is written as $S=S(\Omega=3)$. In addition, if we add a particle with zero internal energy into the system, the possible microstates are given by $\{(0,0,2\epsilon),(0,2\epsilon,0),(2\epsilon,0,0),(\epsilon,\epsilon,0),(0,\epsilon,\epsilon),(\epsilon,0,\epsilon)\}$, yielding more entropy $S'=S'(\Omega=6)$. At the thermodynamic equilibrium, to keep the entropy constant after adding the particle, the additional particle should have the energy $-\epsilon$, resulting in the possible microstates $\{(\epsilon,0,0),(0,\epsilon,0),(0,0,\epsilon)\}$. Hence, the total internal energy decreases from $E=2\epsilon$ to $E'=\epsilon$. In other words, the change in the internal energy is $\Delta E=-\epsilon$. Using the first law of thermodynamics, the chemical potential of the system can be obtained as ${\mu=\big(\partial E/\partial N}\big)_{S, V}=-\epsilon$. In conclusion, for the classical particles with fixing the entropy and the volume constant, the internal energy of the system must decrease when the number of particles increases. It then indicates that the chemical potential in this scenario is always negative. According to the similar behaviors between the aforementioned potentials, the gas system in the presence of a gravitational field might be suitably studied (in the next section) by interpreting the chemical potential as the gravitational potential $\mu(r)\propto\Phi(r)$. Consequently, from Eq.~\eqref{q grad r}, the gradient of the temperature $T_q$ is proportional to that of $\Phi(r)$.
\par
Turning our attention back to another implication of the $q$-Boltzmann transport equation, let us contract Eq.~\eqref{constraint q} with $v^\alpha$, and adopt the same setting as previously done.
As a result, one obtains
\begin{eqnarray}
\displaystyle
-\frac{(1-q)\mu}{T_q^3}\partial_\alpha T_q+\frac{2(1-q)}{T_q^2}\partial_\alpha\mu\hspace{-4cm}&&\nonumber\\
&=&\frac{1}{T_q^2}\partial_\alpha T_q-\frac{1}{T_q^2}\left[1-(1-q)\frac{\mu}{T_q}\right] v_\alpha v^\beta\partial_\beta T_q +\frac{1}{T_q}\left[1-(1-q)\frac{\mu}{T_q}\right]v^\beta\nabla_\beta v_\alpha.
\end{eqnarray}
Employing the relation between the gradients of temperature and chemical potential in Eq.~\eqref{grad T q}, the parallel transport of the four-velocity is expressed as
\begin{eqnarray}
\displaystyle
v^\beta\nabla_\beta v_\alpha 
&=&-\left[\frac{1+(1-q)\displaystyle\frac{\mu}{T_q}}{1-(1-q)\displaystyle\frac{\mu}{T_q}}\right]\frac{1}{T_q}\partial_\alpha T_q\nonumber\\
&& +(1-q)\left[\frac{\displaystyle\frac{2}{T_q}\partial_\alpha\mu+\left\{1-(1-q)\displaystyle\frac{\mu}{T_q}\right\}\frac{1}{T_q} v_\alpha v^\beta\partial_\beta\mu}{1-(1-q)\displaystyle\frac{\mu}{T_q}}\right].
\end{eqnarray}
It is complicated and not easy to analyze. 
In order to capture the physical meaning of this equation, let us expand it around $q=1$ as
\begin{equation}
\displaystyle	
v^\beta\nabla_\beta v_\alpha 
\approx-\frac{1}{T_q}\partial_\alpha T_q+\frac{(1-q)}{T_q^2}\Big(2T_q\partial_\alpha\mu-2\mu\partial_\alpha T_q+T_qv_\alpha v^\beta\partial_\beta\mu\Big).
\label{q-geodesic}
\end{equation}
The gas particles described by Tsallis statistical mechanics cannot move along the geodesic since the right-hand side does not vanish.
This is a key emergent feature when the non-extensive effect is taken into account.
In addition, the particles are not enforced only by the external force due to the temperature gradient, but there is also that from the non-extensivity.
The compact object might be formed undergoing the balance between these forces and gravitational collapse at the thermodynamic equilibrium.
\par
Apart from the classical gas, let us consider the quantum gas in which the gas particles are treated as indistinguishable. By using the same ideas as the quantum gas mentioned in GB statistical mechanics, at the thermodynamic equilibrium, the summational invariant can be written in the suitable form as \cite{Santos:2017gzy}
\begin{eqnarray}
\displaystyle
\left(\frac{\rho'^{(0)}_{q*, (Q)}}{\hat{\rho}'^{(0)}_{q*, (Q)}}\right)^{1-q} \ln_{q} \left(\frac{\rho'^{(0)}_{q*, (Q)}}{\hat{\rho}'^{(0)}_{q*, (Q)}}\right) + \left(\frac{\rho'^{(0)}_{q, (Q)}}{\hat{\rho}'^{(0)}_{q, (Q)}}\right)^{1-q} \ln_{q} \left(\frac{\rho'^{(0)}_{q, (Q)}}{\hat{\rho}'^{(0)}_{q, (Q)}}\right) \hspace{-8cm}&&\nonumber\\
&=&
\displaystyle
\left(\frac{\rho^{(0)}_{q*, (Q)}}{\hat{\rho}^{(0)}_{q*, (Q)}}\right)^{1-q} \ln_{q} \left(\frac{\rho^{(0)}_{q*, (Q)}}{\hat{\rho}^{(0)}_{q*, (Q)}}\right) + \left(\frac{\rho^{(0)}_{q, (Q)}}{\hat{\rho}^{(0)}_{q, (Q)}}\right)^{1-q} \ln_{q} \left(\frac{\rho^{(0)}_{q, (Q)}}{\hat{\rho}^{(0)}_{q, (Q)}}\right),
\end{eqnarray}
where $\hat{\rho}_{q, (Q)} = (1 + \xi \rho_{q, (Q)})$. Consequently, the $q$-equilibrium quantum distribution function is given by
\begin{equation}
\displaystyle
\rho^{(0)}_{q, (Q)}(x^\alpha, P^\alpha)
=\frac{1}{\left[1-(1-q)\left(\displaystyle{\frac{\mu-v_\beta P^\beta}{T_{q,(Q)}}}\right)\right]^{1/(1-q)}-\xi}.
\end{equation}
Substituting it into the Boltzmann equation~\eqref{Boltz eq q} and following the same steps of calculation, the conditions for quantum gas are obtained as follows:
\begin{eqnarray}
\displaystyle
v^{\alpha}\partial_{\alpha}T_{q, (Q)}&=&0,
\label{quan temps grad}\\
v^{\alpha} \nabla_{\alpha} v_{\gamma} &=& -\frac{1}{T_{q, (Q)}} \partial_{\gamma} T_{q, (Q)}.
\label{geo quan Tsallis}
\end{eqnarray}
Surprisingly, the expressions in Eqs. (\ref{quan temps grad}) and (\ref{geo quan Tsallis}) are, respectively, the same as those in Eqs. (\ref{grad T quant}) and (\ref{geodesic like quantum}) described by GB statistical mechanics.
\par
For a static and spherically symmetric system of classical gas, the non-trivial conditions of Eqs. 
(\ref{grad T q}) and (\ref{q-geodesic}) can be written, respectively, as follows:
\begin{eqnarray}
\displaystyle
\partial_{r} T_{q}(r) &=& (1-q) \partial_{r} \mu(r),
\label{q grad r}\\
\displaystyle
v^{\beta} \nabla_{\beta} v_{r} &\approx& (1-q) \left[-\frac{1}{T_{q}} \partial_{r} \mu + \frac{1}{T_{q}^{2}} (2T_{q} \partial_{r} \mu + T_{q} v_{r} v^{\beta} \partial_{\beta} \mu)\right] + \mathcal{O}[(1-q)^{2}].
\label{q-geo r}
\end{eqnarray}
From Eq. (\ref{q grad r}), it is found that the classical gas system with $q \neq 1$ can be non-isothermal. In addition, it is possible to obtain such a system being in a stable configuration due to the effect of the non-extensivity inferred from Eq. (\ref{q-geo r}). 
However, for the quantum gas system, the corresponding non-trivial conditions:
\begin{equation}
\displaystyle
\partial_{r} T_{q, (Q)} = 0, \qquad v^{\alpha} \nabla_{\alpha} v_{r} = 0,
\end{equation}
provide that the system is always isothermal and is in an unstable configuration.
\par
In this section, we have shown that the system of the classical gas with static and spherical symmetry can be non-isothermal and be in a stable configuration, while one of the quantum gas cannot. As aforementioned features, the system of the classical gas, associated with Tsallis statistical mechanics, is suitable for modeling the thermal system of the static and spherically symmetric black hole with multiple horizons. In the next section, we will investigate how the temperatures of the black hole with multiple horizons are generally different and then the systems are not in equilibrium based on GB statistical mechanics. In order to obtain the stable configuration, we will assume that the gas filled in between the horizons is described by Tsallis kinetic theory.

\section{The thermodynamic properties of the black holes}\label{sec:bh}
\par
A black hole can emit the radiation known as the Hawking radiation. It behaves like the black body and carries thermodynamic states such as temperature and entropy. As a result, we can treat the black hole as a thermal object. Note that the thermodynamic quantities of the black hole are evaluated at the event horizons. The black hole with multiple horizons corresponds to multiple thermal objects with generically different temperatures. Therefore, it seems like the thermal equilibrium thermodynamics used to describe the thermal behavior of the black hole cannot be fully utilized for such a black hole. 
\par
In this section, we assume that there exists the gas between the horizons obeying Tsallis statistical mechanics.
For concreteness, we consider the Schwarzschild-de Sitter (Sch-dS) black hole which is the static and spherical solution of the Einstein field equation with a positive cosmological constant, $\Lambda > 0$. The solution can be written as \cite{tangherlini1963schwarzschild}
\begin{eqnarray}
\dd s^2 = -f(r) \dd t^2 + f^{-1}(r) \dd r^2 + r^2 \dd\Omega^2, \,\,\text{with} \,\,
f(r) = 1 - \frac{2M}{r} - \frac{\Lambda}{3} r^{2}, \label{Sch-dS} 
\end{eqnarray}
where $M$ and $\Lambda$ are, respectively, the mass parameter and the cosmological constant. Note that we restrict our consideration to the case $\Lambda > 0$ since it is possible to deal with the black hole with two horizons. The horizons of the spherically static black hole can be obtained by setting $f(r) = 0$. From  Eq.~(\ref{Sch-dS}), it is found that $f(r) \rightarrow -\infty$ for $r \rightarrow 0^{+}$, while $f(r) \rightarrow -\infty$ for $r \rightarrow +\infty$. The horizon function $f(r)$ is concave, and the maximum point can be obtained by solving $\dd f/\dd r|_{r=r_{ex}} = 0$. As a result, one obtains the horizon radius for which the horizon function is maximum as
\begin{equation}
\displaystyle
r_{ex} = \left(\frac{3M}{\Lambda}\right)^{1/3}. \label{rex}
\end{equation}
Substituting Eq.~(\ref{rex}) into Eq.~(\ref{Sch-dS}), the extremum value of the horizon function can be obtained as
\begin{equation}
\displaystyle
f(r_{ex}) = 1 - \big(9\Lambda M^{2}\big)^{1/3}.
\end{equation}
By requiring $f(r_{ex}) \geq 0$, the condition for having the horizon(s) can be constrained as
\begin{figure}[h]\centering
\includegraphics[width=6cm]{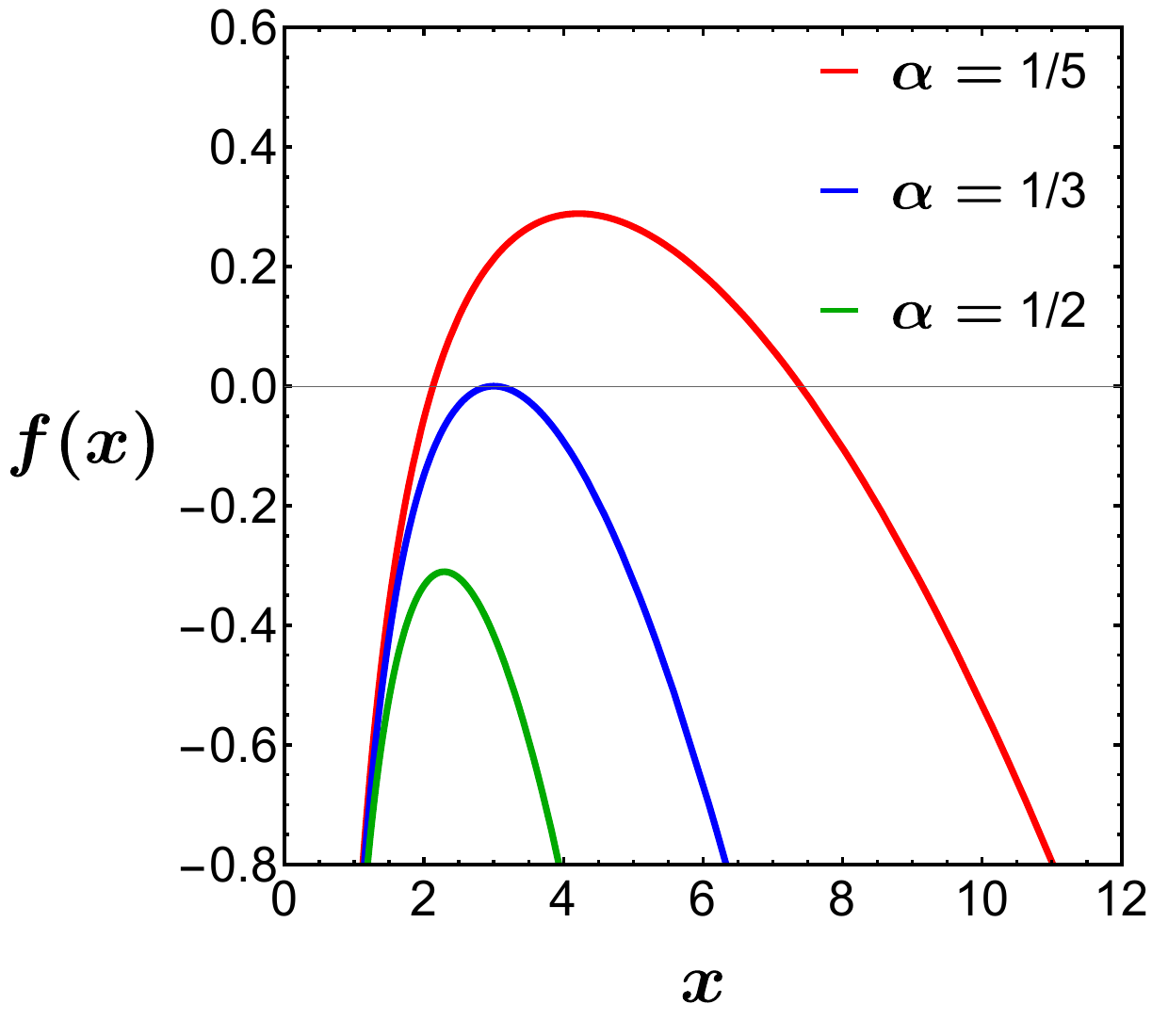}
\caption{Behaviors of horizon function $f(x)$ for Sch-dS black hole versus $x$ by fixing $\alpha = 1/5$ (red), $\alpha = 1/3$ (blue) and $\alpha = 1/2$ (green).} \label{dS horizon}
\end{figure}
\begin{equation}
\displaystyle
0 < \alpha \equiv \sqrt{\Lambda} M \leq \frac{1}{3}. \label{horizon having}
\end{equation}
For convenience, let us define the dimensionless horizon radius as
\begin{equation}
\displaystyle
x = \frac{r}{M}. \label{dimensionless}
\end{equation}
By the definitions in Eqs.~(\ref{horizon having}) and (\ref{dimensionless}), the horizon function in Eq.~(\ref{Sch-dS}) can be rewritten as
\begin{equation}
\displaystyle
f(x) = 1 - \frac{2}{x} - \frac{\alpha^{2}}{3} x^{2},
\end{equation}
The horizon function can be illustrated in Fig.~\ref{dS horizon}, from which there exist two horizons, namely, the black hole horizon; $r_{b}$ and the cosmic horizon; $r_{c}$, where $0 < r_{b} \leq r_{c}$. The horizons can be obtained by solving $f(x_{b}) = 0$ and $f(x_{c}) = 0$. However, the full expressions of the horizons are lengthy and then we do not show them here. However, the approximation for the small value of $\alpha$, $|\alpha| \ll 1$, can be expressed as
\begin{equation}
\displaystyle
\begin{array}{lcr}
\displaystyle x_{b, \text{approx}} = \frac{2}{3} \Big[3 + 4\alpha^{2}\big(1 + 4\alpha^{2}\big)\Big], &\text{and}& x_{c, \text{approx}} = -1 + \displaystyle \sqrt{3}\left(\frac{1}{\alpha} - \frac{\alpha}{2}\right).
\end{array}
\label{horizons}
\end{equation}
\par
As we have discussed, in order to obtain the notion of the hydrostatic equilibrium, we suppose that the gas is in a spherical shell between the black hole horizon and the cosmic horizon. To satisfy the static and spherically symmetric spacetime, the temperature gradient for the non-extensive system can be governed by Eq. (\ref{q grad r}). The difference in the temperature between the black hole horizon and the cosmic horizon can be obtained by integrating Eq.~(\ref{q grad r}) with respect to $r$ from $r_{b}$ to $r_{c}$ as
\begin{eqnarray}
\displaystyle
T_q({r_{c}}) - T_q({r_{b}}) &=& (1-q) \big[\mu (r_{c}) - \mu (r_{b})\big]. \label{change of temp 1}
\end{eqnarray}
In order to obtain the thermal equilibrium between gases and the thermal system of the black hole at the horizon, we suppose that the temperature of the gas system at each horizon is equal to the thermal system of the black hole. From the above equation, one can see that for $q=1$ corresponding to Boltzmann gases, the thermal system of the black hole at both horizons will be in thermal equilibrium. This situation cannot satisfy the black hole thermodynamics in which the temperature at both horizons is not generically equal. For non-extensive gases, $q \neq 1$, it is possible to obtain the thermal systems which are different temperatures. In this case, we can naturally interpret the gases between the horizon being in the hydrostatic equilibrium or the thermodynamic $q$-equilibrium.
\par
Moreover, for  $q\neq 1$, one can see that the temperature gradient is related to the chemical potential gradient of the particles. Therefore, it is worthwhile to discuss how the chemical potential is related to the properties of the black hole. At the extremal case, $r_{b} = r_{c}$, we see from Eq.~(\ref{change of temp 1}) that the difference of the temperature evaluated at each horizon vanishes. In this scenario, the change of the surface gravity also vanishes,  i.e., $\kappa_{b} = \kappa_{c}$. Moreover, the temperature of the black hole evaluated at the horizon is proportional to the surface gravity and we also set the temperature of the gas at the horizon to be equal to one of the black holes. Furthermore, as we have discussed in Sec.~\ref{sec:bh}, at Newtonian gravity, the chemical potential is proportional to the gravitational potential, $\mu (r) \propto \Phi (r)$, in such a way that the tendency of particles to flow from larger radius to the smaller one. Note that the surface gravity can be interpreted as the force per unit mass to hold the particle at the horizon from the observer far away \cite{Poisson:2009pwt}
\begin{equation}
\displaystyle
\kappa_{bh} \equiv a(r_{bh}) = \frac{1}{2} \left|\frac{\dd}{\dd r} f(r_{bh})\right|.
\end{equation}
As a result, it is equivalent to the gravitational potential in most situations. From these results, we can interpret the chemical potential to have a similar form of the surface gravity but now does not need to be evaluated at the horizon. From this argument, we can write the chemical potential in the following form
\begin{equation}
\displaystyle 
\mu (r) = -\mathcal
G \left|\frac{\dd}{\dd r} f(r)\right|,
\label{mu assumption}
\end{equation}
where $\mathcal{G}$ is the positive proportional constant. Note that the negative sign is adopted from the fact that the chemical potential of the classical gas is always negative and the surface gravity is always positive. Substituting Eq.~(\ref{Sch-dS}) into Eq.~(\ref{mu assumption}), one obtains
\begin{eqnarray}
\displaystyle
\mu (r) &=& -\mathcal{G} \left|\frac{2M}{r^{2}} - \frac{2}{3} \Lambda r\right|.
\end{eqnarray}
Let us define a dimensionless chemical potential in terms of the dimensionless variable as
\begin{equation}
\displaystyle
\overline{\mu} (x) \equiv \frac{1}{\sqrt{\Lambda}} \mu (x) = -\frac{2 \mathcal{G}}{3} \left|\alpha x - \frac{3}{\alpha x^{2}}\right|. \label{chemical dS 1}
\end{equation}
\par
The behavior of the temperature between the black hole horizon and the cosmic horizon can be analyzed by using Eq.~(\ref{q grad r}). Integrating both sides of Eq.~(\ref{q grad r}) with respect to $r$, it becomes
\begin{equation}
\displaystyle
T_q(r) = (1-q) \mu(r) + \mathcal{C}, \label{temperature 1}
\end{equation}
where $\mathcal{C}$ is an integrating constant. With GB statistical mechanics, i.e., $q \rightarrow 1$, a black hole and a system of gas are at thermal equilibrium with the same temperature. Immediately, at the limit $q \rightarrow 1$ and at the black hole horizon; $r = r_{b}$, the integrating constant can be obtained as
\begin{equation}
\displaystyle
\mathcal{C}=
\lim_{q \rightarrow 1} T_q(r_{b}) = \frac{1}{4 \pi} \left|\frac{\dd}{\dd r} f(r_{b}) \right|. \label{temperature 2}
\end{equation}
Substituting Eqs.~(\ref{mu assumption}) and (\ref{temperature 2}) into Eq.~(\ref{temperature 1}), the $q$-temperature can be obtained as follows:
\begin{equation}
\displaystyle
T_q(r) = \lambda \left|\frac{\dd}{\dd r} f(r)\right| + \frac{1}{4\pi} \frac{\dd}{\dd r} f(r_{b}),
\label{temperature 3}
\end{equation}
where $\lambda \equiv -(1-q) \mathcal{G}$. Note that, from Fig.~\ref{dS horizon}, the slope of the horizon function at $r_{b}$ is always positive, and hence we have $\displaystyle \left|\frac{\dd}{\dd r} f(r_{b})\right| = \frac{\dd}{\dd r} f(r_{b})$. Substituting the horizon function in Eq.~(\ref{Sch-dS}) into Eq.~(\ref{temperature 3}), the $q$-temperature and its dimensionless version (using $x_{b}$ from Eq.~(\ref{horizons})) can be, respectively, expressed as follows
\begin{eqnarray}
\displaystyle
T_q(r) &=& \lambda \left|\frac{2M}{r^{2}} - \frac{2}{3} \Lambda r\right| + \frac{1}{4\pi} \left(\frac{2M}{r^{2}_{b}} - \frac{2}{3} \Lambda r_{b}\right),\\
\displaystyle
\overline{T}_q(x) &\equiv& \frac{1}{\sqrt{\Lambda}} T (x) = \frac{1}{24\alpha} \left[\frac{81 - 8\alpha^{2} \Big\{3 + 4\alpha^{2} \big(1 + 4\alpha^{2}\big)\Big\}^{3}}{3\pi \Big\{3 + 4\alpha^{2} \big(1 + 4\alpha^{2}\big)\Big\}^{2}} + 16 \lambda \left|\alpha^{2} x - \frac{3}{x^{2}}\right|\right].
\label{T1}
\end{eqnarray}
\begin{figure}[h]\centering
\includegraphics[width=6.5cm]{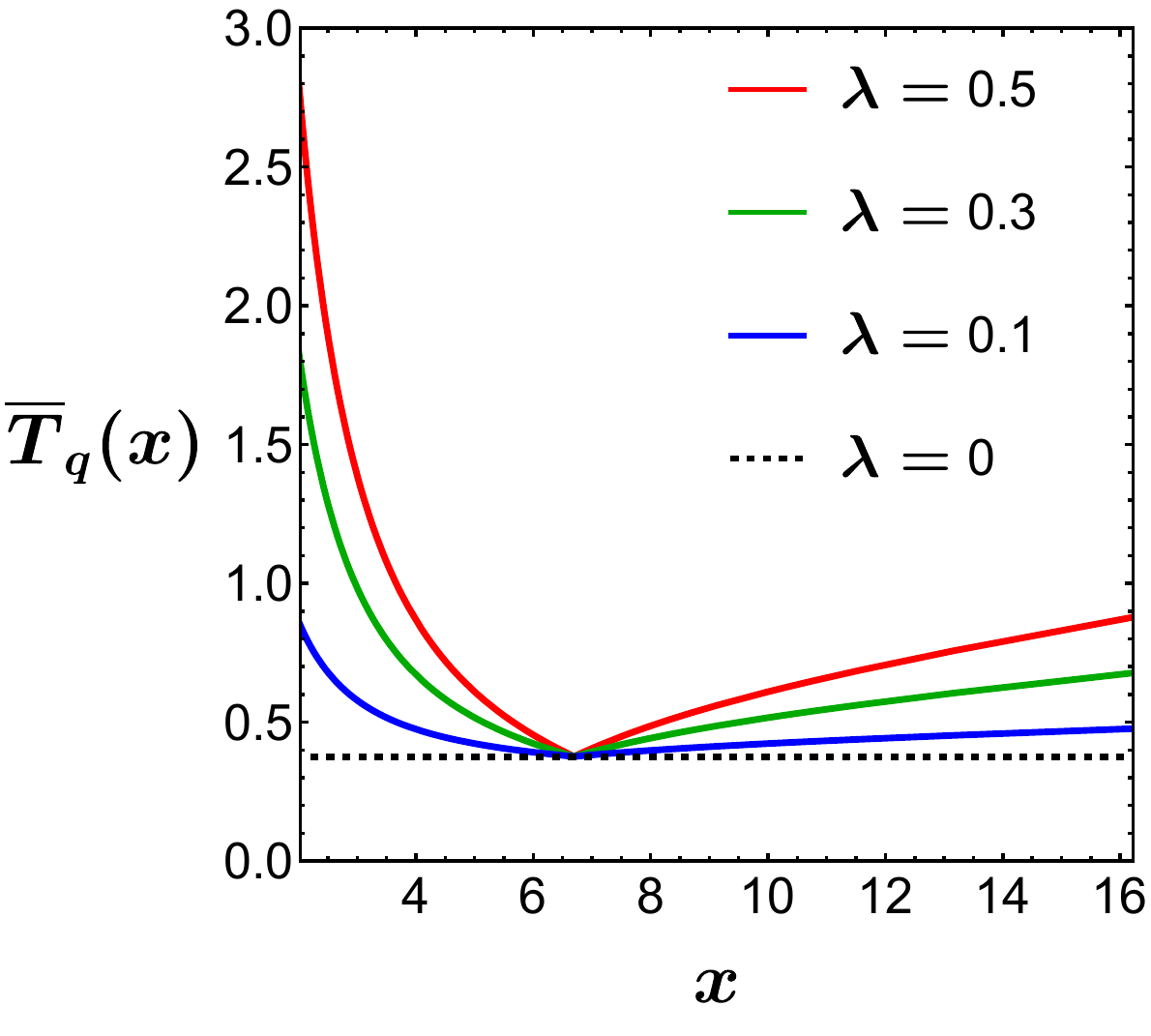}
\hspace{2cm}
\includegraphics[width=5.2cm]{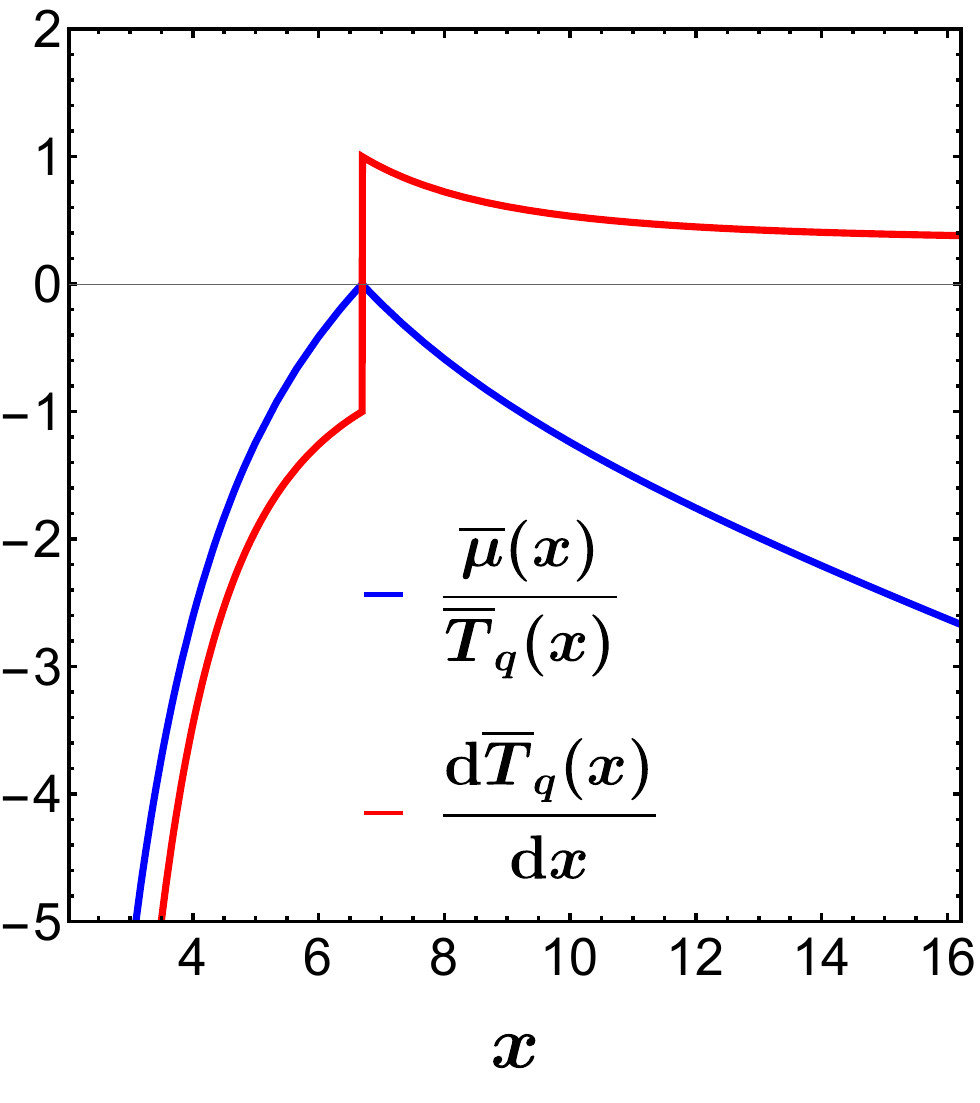}
\caption{The left figure shows the temperature of the system evolving in between $x_{b}$ and $x_{c}$ with the various values of $\lambda$ by fixing $\alpha = 1/10$. The right figure shows the ratio of the chemical potential to the temperature, and the temperature gradient by fixing $\mathcal{G} = 1$, $\alpha = 1/10$, and $\lambda=1/10$.} 
\label{TdS}
\end{figure}
The behavior of the temperature evolving between the black hole horizon and the cosmic horizon can be illustrated on the left-hand side in Fig.~\ref{TdS}. The temperature changes with $x$ between the horizons, causing the temperature gradient between the horizons as seen on the right-hand side in Fig.~\ref{TdS}. Accordingly, the system is generally out of the thermodynamic equilibrium and non-isothermal. Interestingly, the temperature evaluated at the black hole horizon with the non-vanishing non-extensive parameter, $\lambda \neq 0$, is greater than one with the vanishing non-extensive parameter, $\lambda = 0$. Between $x_{b}$ and $x_{c}$, there is a point, $x_{ex}$ obtained from $\dd f/\dd x|_{x=x_{ex}} = 0$, at which the temperature evaluated at the black hole horizons with $\lambda \neq 0$ is equal to one with $\lambda = 0$. The mentioned temperature is written as
\begin{equation}
\displaystyle
\overline{T}_{\text{H}(b)} = \frac{81 - 8\alpha^{2} \Big[3 + 4\alpha^{2} \big(1 + 4\alpha^{2}\big)\Big]^{3}}{72\pi \alpha \Big[3 + 4\alpha^{2} \big(1 + 4\alpha^{2}\big)\Big]^{2}},
\end{equation}
where $\overline{T}_{\text{H}(b)}$ is the Hawking temperature evaluated at the black hole horizon. Although the temperature evaluated at the black hole horizon with $\lambda \neq 0$ is generally greater than one with $\lambda = 0$, at the point $x_{ex}$, these are equal. It is important to note that $\dd f/\dd x|_{x=x_{ex}} = 0$. 
This means that the horizon function $f$ is nearly constant around the position $x_{ex}$, and it can be chosen as unity by re-scaling $x$. In other words, this position may be regarded as an asymptotically flat region.
It is worthwhile to interpret that the observer in this region may measure the temperature of the gas system as the Hawking temperature $\overline{T}_{\text{H}(b)}$. 
According to the mentioned arguments, we can imagine that if there is radiation emitted from the black hole to the observer at rest in flat spacetime, $x_{ex}$, the observer can measure its temperature as $\overline{T}_{\text{H}(b)}$.
\par
It is important to note that the behavior of the gases can be considered by using the ratio of the chemical potential and temperature, $\mu/T$. The gas will behave as the classical gas where the ratio is much more negative $\mu/T \ll 0$ and behave as the quantum gas with $\mu/T \rightarrow 0$. From the right-hand side in Fig.~\ref{TdS}, we see that, at the horizons, the ratio is more negative, compared to other points. On the other hand, the ratio approaches zero at the low temperature or around flat spacetime, i.e., $x_{ex}$. As a result, around flat spacetime or $x_{ex}$, the description of classical gas cannot be fully used since $\mu/T \rightarrow 0$. We need the quantum description to characterize the behavior of the gas. As we have discussed in Sec.~\ref{sec:bh}, for ideal quantum gases, there is no temperature gradient. In other words, the temperature of the quantum gases is always constant, and the system is at the thermodynamic equilibrium. Note that, there are cusps of the temperature profile at $x_{ex}$ illustrated from the left-hand side in Fig.~\ref{TdS}. Then it seems like the temperature is not smooth at this point. However, such a profile at this point is not fully valid since it is the profile of classical gas. In fact, it is to be replaced by the profile temperature of the quantum gas.  
\par
It is worthwhile to note that, even if we try to explain the thermal objects from the multi-horizon black hole, our notion can be applied to the black hole with one horizon such as the Sch black hole. For example, at the limit $x_{c} \rightarrow \infty$ (or $\alpha \rightarrow 0$), the Sch-dS case can reduce the Sch case. The point $x_{ex}$ also approaches infinity, meaning that the temperature of the system with $\lambda \neq 0$ comes closely to the temperature with $\lambda = 0$ at the asymptotically flat spacetime, $x_{ex}$. Therefore, in this case, the observer stays at rest far away from the black hole and will observe the quantum nature of the gases and the temperature is the same as Hawking's temperature of the black hole.
\par
\begin{figure}[h]\centering
\includegraphics[width=6cm]{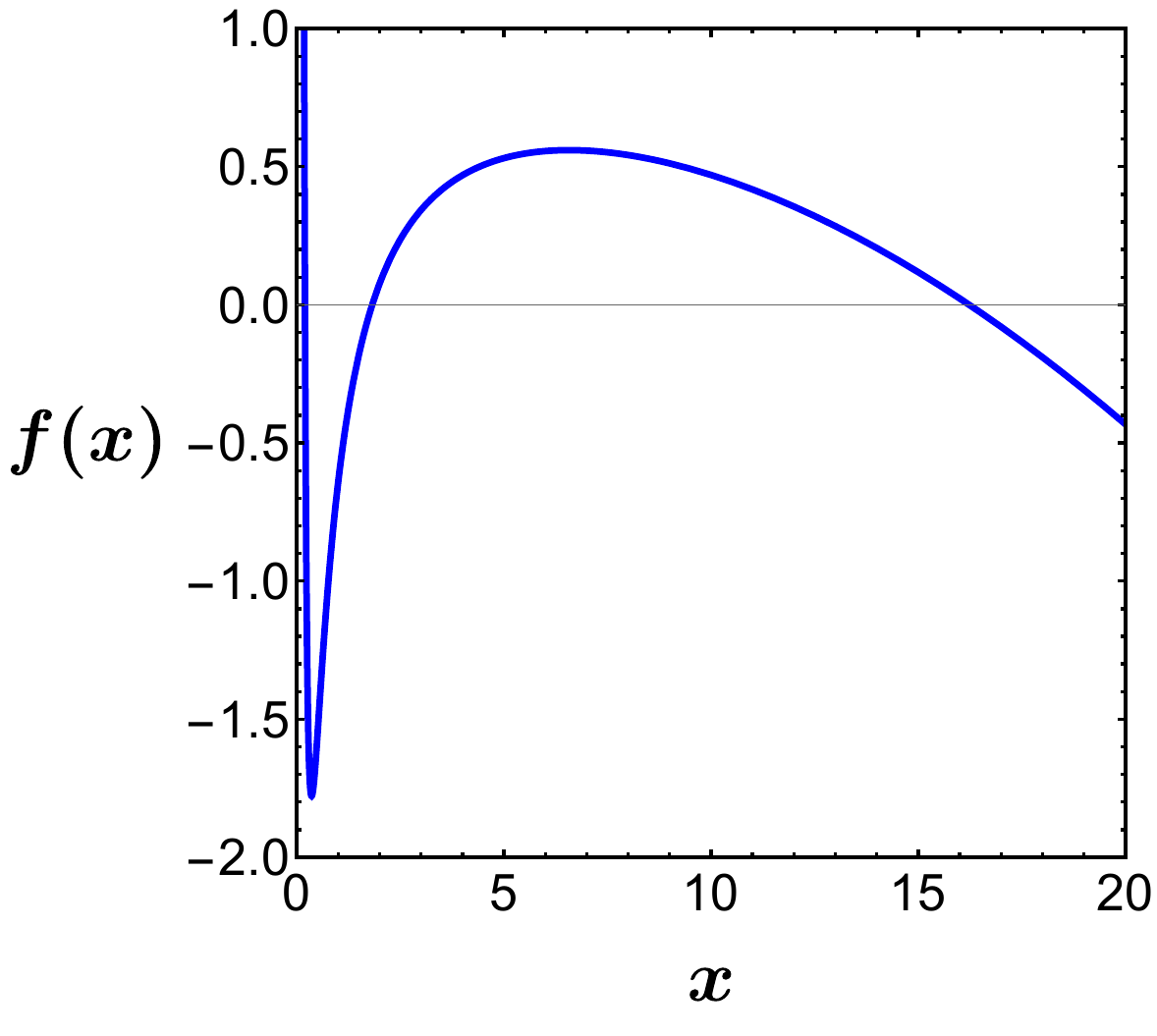}
\caption{Behaviors of horizon function $f(x)$ versus $x$ by fixing $\alpha = 0.1$ and $\beta = 0.6$.} \label{dS horizon-charge}
\end{figure}

Furthermore, we can apply this notion to a black hole with an inner horizon such as a charged black hole. The investigation can describe how the temperature of the gases inside the black hole evolves with $r$. Note that we realize that it may be not possible to access in formation of the matter inside the black hole and the gases must move into the the singularity. However, this investigation will explore how the gases inside the black hole can behave if we require that thermal objects evaluated at the horizons must be in hydrostatic equilibrium. Generally, let us consider the Reissner-Nordstr\"om-de Sitter (RN-dS) black hole where the charge of this black hole is denoted by $Q$. As a result, the horizon function for such a black hole is given by
\begin{equation}
\displaystyle
f(x) = 1 - \frac{2}{x} - \frac{\alpha^{2}}{3} x^{2} + \frac{\beta^2}{x^{2}}, \label{charge 1}
\end{equation}
where the new variables are set as follows $x = r/M$, $\alpha = M \sqrt{\Lambda}$, and $\beta = Q/M$. The behavior of the horizon function is illustrated in Fig.~\ref{dS horizon-charge}. From this figure, we see that, in the range of $x \geq 0$, there exist three horizons, namely, the inner horizon; $x_{in}$, the black hole horizon; $x_{b}$ and the cosmic horizon; $x_{c}$, where $x_{in} \leq x_{b} \leq x_{c}$.
\par
\begin{figure}[h]\centering
\includegraphics[width=6cm]{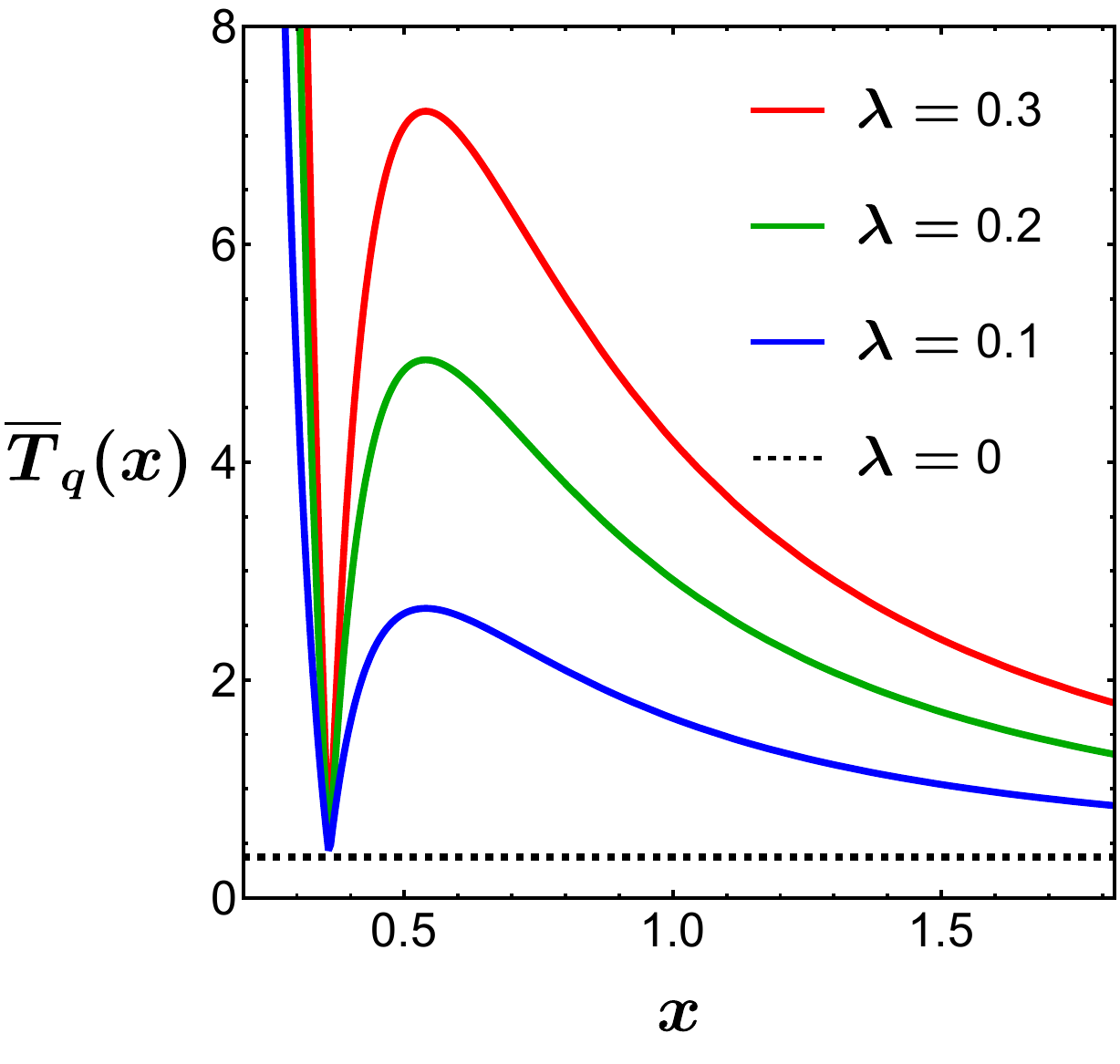}
\hspace{2cm}
\includegraphics[width=6.3cm]{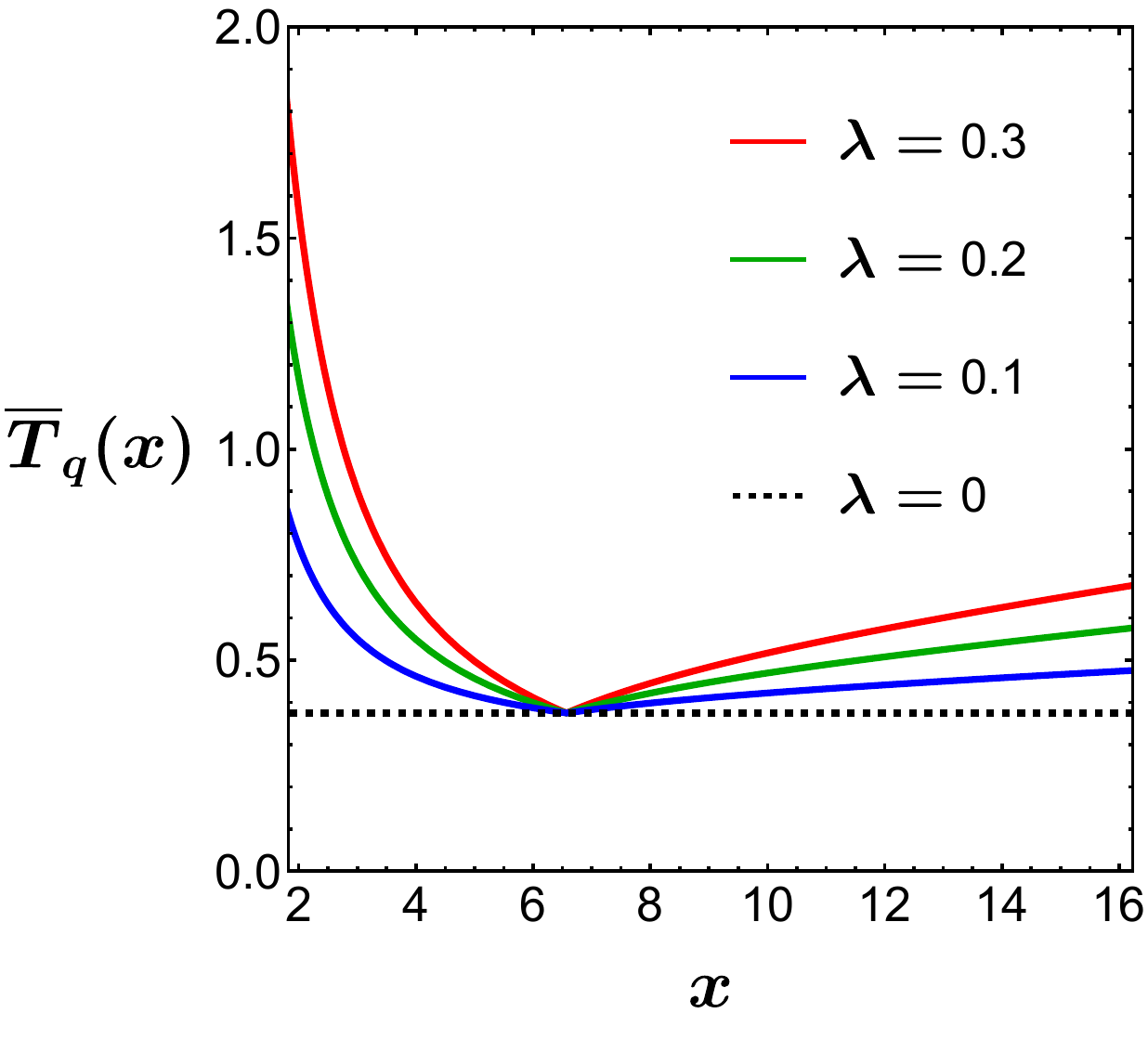}
\caption{The left (right) figure shows the temperature of the system evaluating in between the horizons $x_{in}$ and $x_{b}$ ($x_{b}$ and $x_{c}$) with the various values of $\lambda$ by fixing $\alpha = 0.1$, and $\beta = 0.6$.} 
\label{Tds-charge}
\end{figure}
\begin{figure}[h]\centering
\includegraphics[width=6cm]{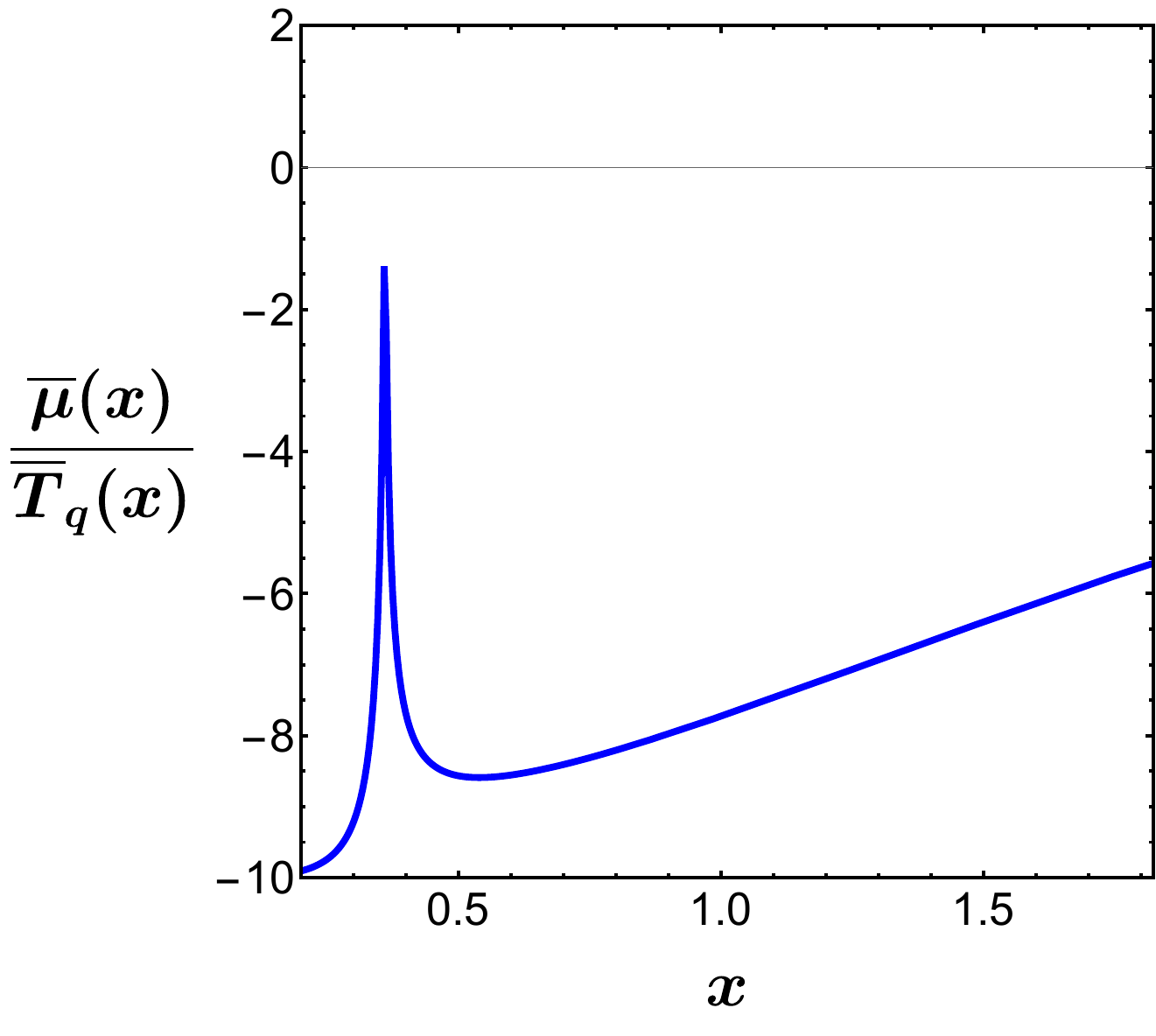}
\hspace{2cm}\quad
\includegraphics[width=6cm]{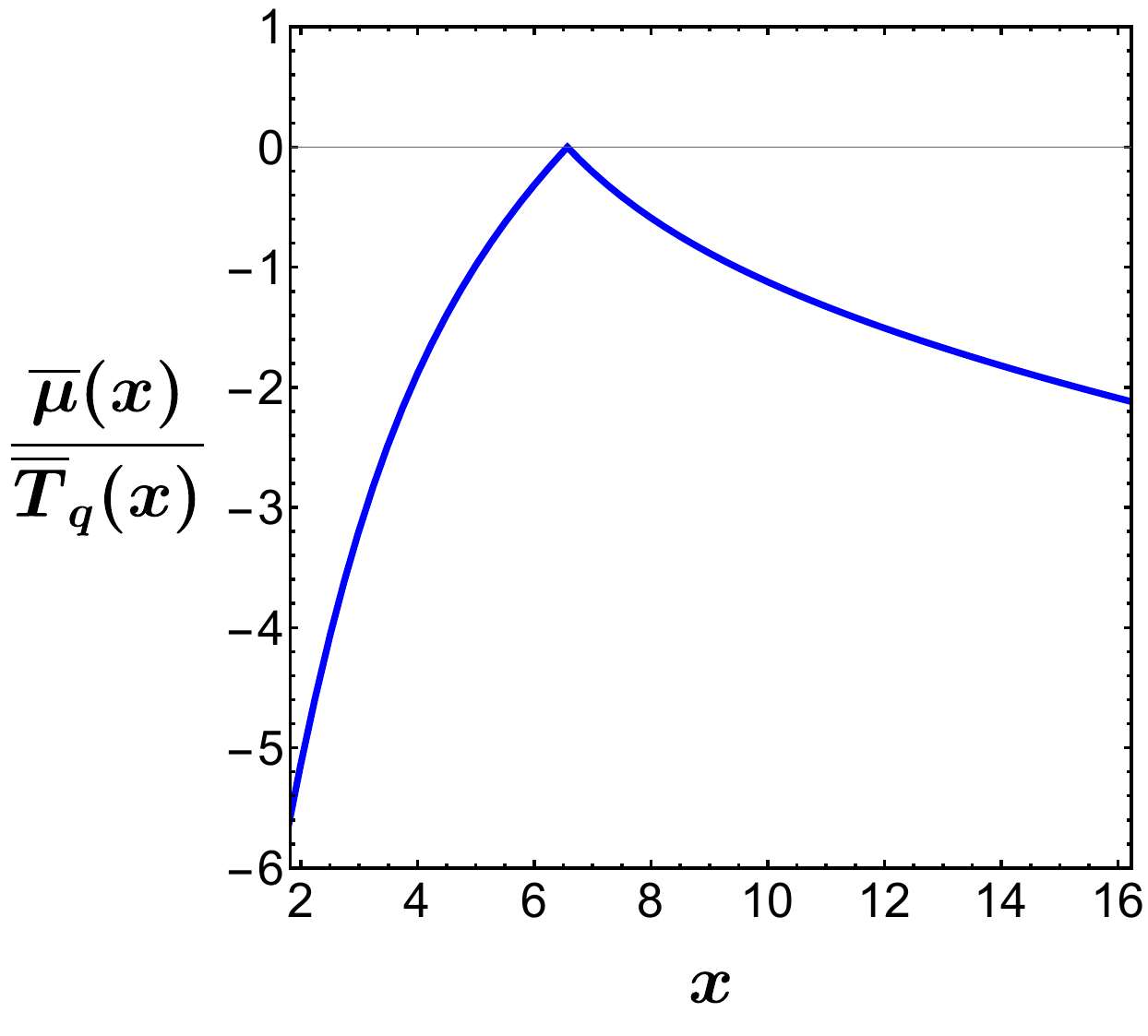}
\caption{The left (right) figure shows the ratio of the chemical potential and the temperature evaluating in between the horizons $x_{in}$ and $x_{b}$ ($x_{b}$ and $x_{c}$) by fixing $\mathcal{G} = 1$, $\alpha = 0.1$, $\beta = 0.6$, and $\lambda = 0.1$.} 
\label{muT-charge}
\end{figure}

By using those thermodynamic quantities, the $q$-temperature and the ratio of the chemical potential and temperature evaluating between the horizon from $x_{in}$ to $x_{c}$ can be obtained. However, their expressions are too lengthy, hence we do not explicitly show them here. However, we confine ourselves to the behavior of the temperature by the plot as illustrated in Fig. \ref{Tds-charge}. The left panel of Fig.~\ref{Tds-charge} shows the temperature profile which changes with $x$ between $x_{in}$ and $x_{b}$, due to the temperature gradient in the black hole's structure. On the right panel of Fig. \ref{Tds-charge}, the temperature profile behaves like those in the case of a Sch-dS black hole as we have discussed. Interestingly, between horizons $x_{in}$ and $x_{b}$, there exist a cusp about $x \approx0.4$, obtained from $\dd f/\dd x|_{x = x_{ex}} = 0$, at which the temperature with $\lambda\neq 0$ approaches one with $\lambda = 0$.
\\
At this point, the ratio of the chemical potential approaches zero as illustrated in the left panel of Fig. \ref{muT-charge}. Therefore, around this point, the gases become quantum gases and the temperature gradient vanishes. This behavior is similar to the one we discussed above. If there exists the flat spacetime ($\dd f/\dd r=0$), the gases behave as the quantum gases. This is also shown in the right panel of the Fig. \ref{muT-charge}.

\section{Conclusion and Discussion}\label{sec:conclude}
\par
Black holes are known to be so strongly gravitating from which even light cannot escape. Classically, they are not treated as thermal objects. However, Hawking demonstrated through quantum field theory in a curved space that black holes can emit thermal radiation and act as the black body~\cite{hawking1975particle}. This paved the way for numerous studies on black hole thermodynamics and allowed physicists to understand such strongly gravitating objects as thermal objects. As the thermal system obtained from the black hole is evaluated at the black hole horizons, the black hole with multiple horizons corresponds to the multiple thermal systems with generically different temperatures. Such thermal systems are out of thermal equilibrium and then the equilibrium thermodynamics may not be in a position to describe such thermal systems. This suggests that other statistical mechanics beyond Gibbs-Boltzmann might be useful to gain insight into understanding such thermal systems. Moreover, the statistical mechanics based on Gibbs-Boltzmann entropy cannot be fully used to explain some systems with long-range interaction while such the system will be compatible with the statistical mechanics based on the non-extensive entropy \cite{antonov1962most, lynden1968gravo, chavanis2002gravitational, Axenides:2012bf, Roupas:2014sda}.  In this work, we used statistical mechanics based on the non-extensive Tsallis entropy to describe how the thermal systems of the multiple horizon black hole can be in the hydrostatic equilibrium instead of thermodynamic equilibrium. 
\par
By assuming that there exist relativistic gases between the black hole horizon and the cosmic horizon, it is possible to investigate the properties of the gases and whether to satisfy the hydrostatic equilibrium or not. 
For GB statistical mechanics, by using the Maxwell-J\"uttner distribution function together with the Boltzmann transport equation, we found that, for the static and spherically symmetric gas system, the temperature gradient vanishes, $\partial_r T=0$  \cite{jiulin2004nonextensive, jiulin2007nonextensivity}. We also found that the self-gravitating system cannot form since gas particles always move along geodesic without counter-balancing terms contributed by the gradient of the temperature. As a result, the gases with GB statistical mechanics do not satisfy the presence of two thermal objects obtained from the horizons of the Sch-dS black hole.
\par
For the Tsallis statistical mechanics, we performed in the same step with GB statistical mechanics but now use the $q$-distribution function and $q$-Boltzmann transport equation instead of the Maxwell-J\"uttner distribution function together with the Boltzmann transport equation, respectively. One of the main results is that there exists the gradient of the temperature which is proportional to the gradient of chemical potential, $v^\alpha\partial_\alpha T = (1-q) v^\alpha\partial_\alpha \mu$. By analyzing the properties of the gravitational potential and the chemical potential, we found that they share the same characteristics. Therefore, the chemical potential is set to be proportional to the gravitational potential to characterize the behavior of the gases between the horizons of the black hole.  Furthermore, we also found that the gas particles do not move along geodesics due to the existence of the temperature gradient and non-extensivity, i.e., $q\neq 1$. This provides us with a possible way to describe two thermal objects of the Sch-dS black hole with the hydrostatic equilibrium. It is worthwhile to note that by using the same step of the calculation, we also found that the quantum relativistic gas does not provide the temperature gradient even if the non-extensivity is taken into account. 
\par
Due to the hydrostatic equilibrium of the non-extensive gas, it is worthwhile to apply this kind of gas to describe how two thermal objects for the multi-horizon black hole can be in equilibrium.  In the present paper, we focus on the static and spherically symmetric black hole with multiple horizons. For the Sch-dS black hole, there exist two horizons denoted by black hole horizons (smaller one) and cosmic horizon (larger one). According to the spherical symmetry, the temperature is dependent only on radial coordinate $r$. Then the equation for the temperature gradient can be written as $\partial_r T = (1-q) \partial_r \mu$. In order to obtain the thermal equilibrium between gas and the thermal system of the black hole at the horizon, we suppose that the temperature of gas at each horizon is equal to the thermal system of the black hole. Then, the temperature of the gas can be written as $\displaystyle{T(r) = T_b + \lambda \frac{\dd f(r)}{\dd r}}$ where $T_b$ is the Hawking temperature at the black hole horizon and $\lambda = -\mathcal{G} (1-q)$, $\mathcal{G}$ is a proportional constant. Now $\lambda$ characterizes the non-extensive effect. The temperature profile with various $\lambda$ can be found on the left panel of Fig. \ref{TdS}. Moreover, by observing the ratio $\mu/T$ on the right panel of Fig. \ref{TdS}, we found that the gas behaves classically near the horizons and acts like quantum gas around flat spacetime. If we observe the gas far away from the black hole we will observe the quantum nature of the gas and their temperature will be constant and equal to one for the Hawking radiation. 
\par
Even though we attempt to describe the thermal properties of black holes with multiple horizons, this notion can be applied in a black hole with one horizon such as Sch black hole. In this case, we found that the gas near the black hole horizon behaves as a classical one, while we will observe the quantum nature of the gas far away from the black hole with the same temperature as one from Hawking radiation. Moreover, we can apply this notion to a black hole with an inner horizon such as a charged black hole. In this case, the gas inside the black hole will be classical and there exists a particular radius at which the gas becomes quantum gas. Even though it may be not possible to access information on the matter inside the black hole and the gas must move into the singularity, this will explore how the gas behaves inside the black hole if we require that thermal objects evaluated at the horizons must be hydrostatic equilibrium. This might pave the way to study the behavior of the matter inside the black hole in terms of black hole thermodynamics. It is worthwhile to emphasize here that even the gas will support the thermal objects from the black hole horizons to be in a stable configuration with hydrostatic equilibrium, their thermal stability themselves should be explored. We will leave this investigation for further work.

\section*{Acknowledgement}
This research has received funding support from the NSRF via the Program Management Unit for Human Resources \& Institutional Development, Research and Innovation [grant number B37G660013].

\bibliography{ref.bib}

\end{document}